\documentclass{elsart}

\usepackage{psfig}

\usepackage{amssymb}
\newcommand{\beq}{\begin{eqnarray}}
\newcommand{\eeq}{\end{eqnarray}}
\newcommand{\punkt}{\quad .}
\newcommand{\komma}{\quad ,}
\newcommand{\romg}{\mbox{\rm g}}
\newcommand{\romf}{\mbox{\rm f}}
\newcommand{\romerf}{\mbox{\rm erf}}
\newcommand{\Tesla}{\mbox{\rm T}}
\newcommand{\HTesla}{\mbox{\rm T/$\mu_0$}}
\newcommand{\mmm}[1]{\mbox{$\langle M \rangle_{#1}$}}
\newcommand{\mean}[1]{\mbox{$\left \langle #1 \right \rangle$}}

\newcommand{\mmq}[1]{\mbox{$\langle M \rangle_{#1}^2$}}

\newcommand{\sigmas}{\sigma_s}
\newcommand{\sigmaf}{\sigma_f}
\newcommand{\sigmafhat}{\hat{\sigma}_f}
\newcommand{\ynorm}{\mbox{$\frac{\bar{H}_s}{\sigma_s}$}}
\newcommand{\norms}{\mbox{$\displaystyle \frac{2}{\sqrt{\pi} \sigmas
\left( 1+{\rm erf} \left( \ynorm \right)\right)}$}}

\newcommand{\Hex}{\mbox{$H_{ext}$}}
\newcommand{\hex}[1]{\mbox{$H_{ext}^{(#1)}$}}
\newcommand{\hiup}[1]{\mbox{$H_{i,\uparrow}^{(#1)}$}}
\newcommand{\hidown}[1]{\mbox{$H_{i,\downarrow}^{(#1)}$}}

\newcommand{\wdd}{\mbox{$w_{\downarrow\downarrow}$}}
\newcommand{\wud}{\mbox{$w_{\uparrow\downarrow}$}}
\newcommand{\wdu}{\mbox{$w_{\downarrow\uparrow}$}}
\newcommand{\wuu}{\mbox{$w_{\uparrow\uparrow}$}}
\begin{document}
\journal{JMMM}
\begin{frontmatter}



\title{Theory of Thermal Remagnetization of Permanent Magnets}


\author[schumann]{R. Schumann\corauthref{cor1}}
\corauth[cor1]{Corresponding Author}
\ead{schumann@theory.phy.tu-dresden.de}
\author[jahn]{L. Jahn }

\address[schumann]{Institute for Theoretical Physics, TU Dresden, D-01062 Dresden, Germany}
\address[jahn]{Institute for Applied Physics, TU Dresden, D-01062 Dresden, Germany}

\begin{abstract}
A self-consistent mean-field theory explaining the thermal 
remagnetization (TR) of polycrystalline permanent magnets is given.  
The influence of the environment of a grain is treated
by an inclusion approximation, relating the field inside the grain to the local
field outside by means of an
internal demagnetization factor $n$.
For the switching fields and the fluctuations of the local fields around the
mean field Gaussian distributions of widths $\sigmas$ 
and $\sigmaf$ resp. are assumed. 
The isothermal hysteresis curve, the recoil curves, and 
the TR in dependence on the model parameters $n$,  $\sigmas$, and $\sigmaf$ are
calculated. 
Furthermore, the influence of the initial temperature and the 
strong dependence of the TR on the demagnetization factor of the sample are studied, 
and it is shown that for reasonable parameter sets TR effects
up to 100 \%
are possible.
The theoretical results correspond well with the experimental situation.
\end{abstract}

\begin{keyword}
Thermal Remagnetization \sep  
Permanent Magnets\sep
Hysteresis\sep 
Inclusion Approximation\sep 
Switching Field Distribution
\PACS 75.50Vv \sep 75.50Ww \sep 75.60Ej
\end{keyword}
\end{frontmatter}
%
%
\section{Introduction}
\label{Introduction}
It is common knowledge that increasing temperature destroys 
ferromagnetism, due to the related decrease of both the 
saturation magnetization and the coercivity. Thus an experiment, 
where an unmagnetized
sample is heated to some hundred degrees and becomes thereby magnetized, 
is a challenge to our understanding of permanent magnetism. 
This effect was discovered in SmCo$_5$ about 25 years ago 
\cite{Lifshits74,Kavalerova75} and named thermal remagnetization (TR). 
The experimental procedure, sketched in Fig. \ref{prinzip}, includes the
saturation and isothermal dc-demagnetization via the points ``1''
 and ``2'' and a subsequent heating.
The largest TR-effect, i.e. more than 80 \%
of the saturation magnetization at the initial temperature, has been observed 
in well-aligned sintered SmCo$_5$ for a closed magnetic circuit
\cite{Kavalerova75,Lileev77}. 
In open circuits the TR depends strongly on the 
demagnetization factor $N$ of the sample \cite{Livingston84,Jahn85a}. 
In well texturized NeFeB magnets 
the TR-effect is lower, but well measurable \cite{Jahn85a,Mueller88}. 
In contrast to the nucleation controled
SmCo$_5$- and NdFeB-magnets the effect is absent \cite{Livingston84,Scholl87} 
or at least very small \cite{Jahn91} in
high coercivity Sm$_2$Co$_{17}$ sintered magnets, 
which are believed to be pinning controled.
In spite of the good experimental situation the theoretical understanding
is dissatisfying. There is agreement on the fact that the average 
coercivities of positively and negatively magnetized grains have to be
different and that the
temperature dependence of the coercivity is essential for
understanding the TR. The latter was demonstrated by experiments on
barium ferrite, where a sign-change of the temperature coefficient of the coercive
field H$_{\rm C}$
results in an ``inverse'' TR effect, i.e. a remagnetization 
upon cooling \cite{Jahn85a}.
Livingston and Martin \cite{Livingston84} argued that the
TR in SmCo$_5$ is due to  the change of the grains from single-domain to multi-domain 
state. Their mechanism can not explain TR to more than 50 \%. 
We proposed an alternative 
mechanism \cite{Jahn85a}, where the TR
is caused by the fluctuating fields in the neighbourhood of ``hard'' grains. 
This first model aimed at a qualitative understanding 
of the TR solely, therefore oversimplified distributions of the switching fields
and the field fluctuations were used. 
In a subsequent calculation \cite{Schumann87} we 
introduced Gaussian distributions both for the switching fields
and for the field fluctuations. Furthermore the distribution width of the 
latter was chosen as function of the average magnetization. 
This brought the resulting TR curves closer to the experimental situation,
especially well below the temperature $\rm T_{max}$, where the maximum 
TR occurs, but was not able to explain the vanishing TR at 
temperatures above. The same kind of distributions for the fluctuation fields
were used by M\"uller et. al. \cite{Mueller88}. They presented
a mean-field theory devoted mainly to the coercivity of
NdFeB, where non-magnetic phases as well as multi-domain magnetic grains 
are present. They also derived a formula for the maximum TR.
Whereas they took the field fluctuations into account while calculating
the demagnetization curve, they neglected these fluctuations in that part of
their paper, which was devoted to the TR. 
The resulting overestimation of the TR for NeFeB they attributed to
that neglection of the field fluctuations, and furthermore to the pinning of 
domain walls, as well as to the influence of the
sample demagnetization factor. 
Supported by the experimental fact,
that SmCo$_5$ of lower to medium coercivity in its dc-demagnetized state contains 
multi-domain grains \cite{Livingston73,Livingston84,Zaytzev88}, 
Lileev et al. \cite{Zaytzev89,Lileev92} presented a theory, 
which takes into account both
multi-domain grains, and field fluctuations caused by the interacting field
of neighbouring grains. Since they included the mutual dipol-dipol
interaction of the grains they had to carry out numerical simulations 
which makes it difficult to handle their theory, especially if one is 
interested in the influence of basic parameters e.g. the distribution width 
of the field fluctuations, the influence of the initial temperature etc..
They calculated a TR curve which has a
rather sharp peak. Whereas the magnitude of the maximum effect is comparable
to the experiment, the shape of the theoretical curve differs from
the experimental situation. Nevertheless, we share the opinion that the
influence of the nearest neighbourhood of a grain is an essential 
for every elaborated theory of the TR. For this task the current paper extends
our former theory \cite{Schumann87} by an inclusion approximation. 
Furthermore we take into account the possibility 
that the grains may nucleate into
multi-domain states. The paper is organized as follows: In the next
section we present the model. Since a theory of the TR presumes a calculation
of both the isothermal hysteresis loops and the recoil curves from every point of
the hysteresis loop, we present the related theory and the results in subsection 
3.1 and 3.2 respectively. The calculation of the TR due to heating is the topic
of subsection 3.3. In the last section we discuss the results.
\section{The model}
\label{themodel}
The polycrystalline permanent magnet is an ensemble of high-coercive magnetic
uniaxial grains.
To fix our model we have to define the properties of both a single grain and 
the ensemble. 
\subsection{Properties of a single grain}
Every grain is characterized by 
\begin{itemize}
\item[(A1)]  $M_S(T)$, the temperature dependent saturation magnetization, 
\item[(A2)]  $H_s(T)$, the temperature dependent switching field. This is the 
 	absolute value of the field needed for reversing the magnetization 
	of the grain in a closed circuit.
\item[(A3)]  $n$, the ``internal'' demagnetization factor, which is determined by
	the shape of the grain,   
\item[(A4)]  $\vec{c}$, the direction of the easy axis due to the high uniaxial 
             anisotropy 
\item[(A5)]  $V$, the volume of the grain.
\end{itemize}
\subsection{Properties of the ensemble} 
Regarding the ensemble we assume
\begin{itemize}     
\item[(B1)] all grains exhibit the same $M_S(T)$, 
\item[(B2)] the easy axis $\vec{c}$ of the grains are completely aligned 
	(ideal texture),
\item[(B3)] the grains differ in their switching fields $H_s$, but the
	  distribution of these switching fields should be known.
	  For simplicity we assume a Gaussian normalized with respect
	  to the region $0<H_s<\infty$,
\beq
\romg(H_s)=\norms \,\, {\rm e}^{
-\left (\frac{H_s-\bar{H}_s}{\sigmas} \right)^2} \komma
\label{gs}
\eeq	
 with the ``mean switching field'' $\bar{H}_s$ and the distribution width $\sigmas$.
\item[(B4)] the temperature dependence (not the absolute value!) of the switching
	fields of all grains is equal.
\end{itemize}
It is problematic to get direct information both on the temperature
dependence of the switching fields and their distribution.
What can be measured is the average magnetization in dependence on the external
magnetic field, the temperature and the time. The switching field distribution 
can not be measured directly. 
In non-interacting  ensembles the switching field distribution may be
determined from remanence measurements \cite{Remanenzmethode}, but 
if the grains interact strongly this method fails. 
From the Henkel plots measured for NdFeB
\cite{Henkelplot_allgemein,HenkelplotNdFeB} 
we know that most sintered hard magnets are strongly interacting ensembles. Thus, for the
moment, we have to regard $\bar{H}_s$ and $\sigmas$ as model parameters. Fortunately
we will see later on that our theory delivers a possibility to relate
$\bar{H}_s$ to the coercivity $H_C$ of the sample.
\subsection{Inclusion approximation}
Since we believe that the TR is caused mainly due to grain interactions
we want to explain the approximations in detail.
On a macroscopic length scale the internal magnetic field and the magnetization of the
sample are homogeneous.
But, if  the length scale is reduced to the order of some grain diameters the
magnetization becomes coarser and the inhomogeneities gain influence.
Averaging over volume elements containing a few grains only, yields values,
which deviate from the average.
Thus, we consider every grain as an inclusion embedded in a local environment, 
which may differ stochastically in its magnetic field and magnetization 
from the related averaged values. We show this schematically in Fig. \ref{inclusion}. 
For simplicity we approximate the magnetization in the vicinity of the
inclusion by the mean magnetization 
$\mean{M}$, neglecting the mentioned fluctuations. 
Otherwise we allow for field fluctuations  $\Delta H = H-\mean{H}$  
around the mean internal field 
in the environment of a grain. The mean internal field is related to
the external applied field $\Hex$ according to 
\beq
\mu_0\mean{H}=\mu_0\Hex-N\mean{M} \komma
\label{extdemag}
\eeq
with $N$ being the demagnetization factor of the sample. 
The field fluctuations are characterized
by the following assumptions 
\begin{itemize}
\item[(C1)] The local magnetic fields (in $\vec{c}$-direction) 
are Gaussian distributed
\beq
\romf(H)&=& \frac{1}{\sqrt{\pi} \sigmafhat} \,\,
{\rm e}^{
-\left(\frac{\Delta H}{\sigmafhat}\right)^2} \qquad \mbox{with} \qquad \Delta H 
= H-\mean{H} \punkt
\label{fh}
\eeq
\item[(C2)] The distribution width $\sigmafhat$ is itself a function of
the mean magnetization. We set
\beq
\sigmafhat &=& \sigmaf  \left ( 1 - \frac{\mean{M}^2}{M_S^2} \right ) \punkt
\label{sigmafvonM}
\eeq
\end{itemize} 	
Assumption C1 is a consequence of the central limit theorem of probability 
theory.
Of course the local fields fluctuate around the mean internal field
also in direction, thus, strictly
speaking eq. (\ref{fh}) accounts for the fluctuations of the z-component only. 
Furthermore the deviation of the field direction from the easy axis gives 
rise to rotation processes, but this
is surely a small effect in ideally aligned magnets with high anisotropy 
constants.
Assumption C2 attempts to take into account that in an ideally textured 
magnet the fluctuation width is zero in the saturated state and maximum
in the dc-demagnetized state. If $\sigmafhat(\mmm{})$ is expanded to second order
and the expansion parameters are fixed  by these
requirements together with the demand that 
$\sigmafhat(\mmm{})$ has to be a symmetric
function one gets eq. (\ref{sigmafvonM}).
\section{Field- and temperature-dependent magnetization}
\label{themagnetization}
\subsection{Calculation of the isothermal demagnetization curve}
\label{isothermalcurve}
After the saturation all grains are up ($\uparrow$) magnetized. They posses
their maximum switching fields and $\mmm{}=M_S$ holds, due to assumption B2.
If the external field is lowered to $\hex{1}$ (point ``1'' in Fig. \ref{prinzip}) 
the magnetization of a grain is switched down ($\downarrow$) if its internal 
field $\hiup{1}$ is lower than $-H_s$ and the internal field 
after switching $\hidown{1}$ will be smaller than $+H_s$.
Thus we have the two conditions
\beq
\hiup{1}&<&-H_s \quad \quad \mbox{(``switching condition'')} \label{schalt1} 
\komma \\ 
\hidown{1}&<&+H_s \quad \quad \mbox{(``not-back-switching condition'') \quad .}  
\label{schalt2}
\eeq
The internal fields before and after switching are with respect to the above
introduced approximations: 
\begin{eqnarray}
\mu_0\hiup{1} &=& \mu_0 H - n ( M_s -\mmm{1}) \komma \\
\mu_0\hidown{1} &=& \mu_0 H - n( - M_s-\mmm{1}) \punkt
\end{eqnarray}
Here $H$ is the local field in the environment of the grain. 
A variation of the external field $\Hex$ results in a change of $\mmm{}$
on the one hand and on the other hand it changes the probability $f(H)dH$ 
that the field in the environment of a grain is between $H$ and $H+dH$ due
to the relation (\ref{extdemag}) and eq. (\ref{fh}).
If the switching condition (\ref{schalt1}) is fulfilled but the condition 
(\ref{schalt2}) is not fulfilled, i.e.
\beq
\hidown{1}&>&H_s \quad \quad \mbox{(``back-switch condition'')} 
\label{schalt3}
\eeq
holds,
the grain cannot jump into a stable state. 
If the grain is large enough
it solves the conflict by an incomplete jump, what turns its single-domain 
state (SDS) into a multi-domain-state (MDS). 
We will call such grains ``weak''.
Whereas the ``hard'' grains can exist in states with $\pm M_S$ only, the
magnetization
of a weak grain $\mmm{i}$ is an average over the upwards and downwards
magnetized volume fractions within the grain. From phase theory of
180$^{\circ}$-domains we get for the averaged magnetization of the
i$\rm ^{th}$ grain in dependence on the local field in the environment
(cf. Appendix A)
\beq
\mmm{i}&=&\mmm+\frac{\mu_0 H}{n} \quad .
\label{weakmagnetization}
\eeq 
Opposite to the hard grains the weak grains have no memory since the
magnetization follows the local field immediately as long as $|\mmm{i}|<M_S$
holds. Otherwise they are saturated up- or downwards.  
If one calculates the probability that a grain with a given $H_s$ is in a MDS
the switching condition (\ref{schalt1}) limits the integration over the 
local field distribution from above,
whereas the back-switching condition (\ref{schalt3}) limits from below. The
related probability is
\beq
p_{w}(H_s)&=& \int\limits_{H_L}^{H_H}\!\!dH\; {\rm f}(H) 
\eeq
The two limiting fields are
\beq
\mu_0H_H&=&-\mu_0H_s+nM_S-n\mmm{} \komma \\
\mu_0H_L&=&\mu_0H_s-nM_S-n\mmm{} \punkt
\eeq
The requirement that the upper limit has to be greater then
the lower one yields the condition
\beq
\mu_0H_s&<&n M_s  \label{Domaenenbedingung} \punkt
\eeq
Thus, grains with switching fields smaller than their own internal
demagnetizing field are weak.
The magnetization of the volume fraction of the weak grains with a given $H_s$ 
is 
\beq
\mmm{w}(H_s) &=& 
M_S \int\limits_{H_H}^{\infty}dH\,  \romf(H)
-M_S \int\limits_{-\infty}^{H_L}dH\,  \romf(H)
+\int\limits_{H_L}^{H_H}dH\, \romf(H) \, \mmm{i}
\punkt
\eeq
The integral may be evaluated yielding
\beq
\mmm{w}(H_s) &=&
-\frac{M_S}{2} \left ( \romerf(y_H)+\romerf(y_L) \right )\\
&& +\, 
\frac{1}{2} \left (\frac{\mu_0\hex{1}}{n}+\mmm{1} \Big(1-\frac{N}{n} \Big) \right )
\left ( \romerf{(y_H)}-\romerf{(y_L)} \right ) \nonumber \\ 
&&-\frac{\mu_0\sigma_f}{2\pi n } \left( {\rm e}^{-y_H^2}-{\rm e}^{-y_L^2} 
\right )  \nonumber 
\label{mvonHsw}
\eeq
with
\beq
y_H&=&\frac{-\mu_0H_s+n M_S-n\mmm{1}-\mu_0\hex{1}+N\mmm{1}}{\mu_0\sigmaf \left (1-\mmq{1}/M_S^2
\right)} \label{y_H}\\
y_L&=&\frac{\mu_0H_s-n M_S-n\mmm{1}-\mu_0\hex{1}+N\mmm{1}}{\mu_0\sigmaf \left (1-\mmq{1}/M_S^2
\right)} \label{y_L}
\punkt
\eeq
The index 1 at $\hex{1}$ and $\mmm{1}$ indicates that the calculation is for
point 1 in Fig. \ref{prinzip}.\\

For hard grains $H_L$ is higher than $H_H$. If these grains fulfill
the switching condition (\ref{schalt1}), condition (\ref{schalt2}) 
will be fulfilled automatically.
The probability to find a grain downwards magnetized if a field \hex{1}
is applied is 
\begin{eqnarray}
p_1(H_s)=\int\limits_{-\infty}^{+\infty}\!\!dH\; {\rm f}_{(1)}(H) \;
\Theta(-H_s-\hiup{1})\;
\Theta(H_s-\hidown{1}) \, ,
\label{p1def}
\end{eqnarray}
with $\hiup{1}$ and $\hidown{1}$ being the fields within the grain in the
upwards and downwards magnetized state resp. and 
$\Theta(x)$ is the Heaviside function. ${\rm f}_{(1)}$ is the field
distribution eq. (\ref{fh}) with $\mu_0\mean{H}=\mu_0\hex{1}-N\mmm{1}$.
The integration in eq. (\ref{p1def}) is easily done yielding
\beq
p_1(H_s)&=&\frac{1}{2}\big (1+ \romerf(x_H) \big) \\
\mbox{with}\qquad&& 
x_H = \frac{-\mu_0 H_s+n M_s -n \mmm{1}-\mu_0\hex{1}+N \mmm{1}}
{\mu_0\sigmaf \left (1-\mmq{1}/M_S^2 \right)} \nonumber
\label{p1}
\eeq
For the magnetization of such a fraction of hard grains with a given $H_s$ we find
\beq
\mmm{1}(H_s)&=& M_S\, (1-2 p_1(H_s))= - M_S \,\romerf(x_H)
\eeq
The total magnetization  of the sample one gets from averaging with 
respect to $H_s$
\beq
\mmm{1}&=&\int\limits_{0}^{nM_S/\mu_0}\,d\,H_s\,\romg(H_s)\, \mmm{w,1}(H_s) +
\int\limits_{nM_S/\mu_0}^{\infty}\,d\,H_s\,\romg(H_s)\, \mmm{1}(H_s) 
\label{demagw}
\eeq
Here the integration has been splitted due to the different contributions of
the weak and hard grains. For the weak grains we have to use 
eq. (\ref{mvonHsw}) with $\Hex$ being $\hex{1}$ and $\mmm{}$ being $\mmm{1}$.
Since the right hand side of eq. (\ref{demagw}) depends on the mean
magnetization $\mmm{1}$ itself, we have to solve this implicit equation
numerically. This is done by fixing $\mmm{1}/M_S$ and searching the related $\hex{1}$.
Fortunately there is always exactly one solution for $-1<\mmm{1}/M_S<1$.
The aim of the present paper is to clarify, how the TR depends on the model
parameters $\sigmas$, $\sigmaf$ as well as on the internal demagnetization factor $n$.
We vary these parameters around the values 
$M_S=1\Tesla$, $\sigmas=1.5\HTesla$, $\sigmaf=0.5\HTesla$, $n=0.333$, and
a mean value of the switching field distribution  
of $\mu_0\bar{H}_s=3.0\Tesla$. The values of $M_S$, $\bar{H_s}$, and $\sigmas$  
resemble SmCo$_5$ (\mbox{VACOMAX 200}), where we measured at T=300K e.g. 
$M_S=0.972\Tesla$,
$\mu_0H_C=2.87\Tesla$. $\sigmas$ is a rough estimate of the width of
the differentiated experimental demagnetization curve.   
The results for the calculated demagnetization curve are plotted in 
Fig. \ref{recoil}, 
where a spherical sample was assumed ($N=1/3$).
\subsection{Calculation of the isothermal recoil curve}
\label{isothermalrecoil}
Next, as shown in Fig. \ref{prinzip} (Point ``1'' to point ``2''), 
the external field is changed from
$\hex{1}$ to $\hex{2}$ with $\hex{2}>\hex{1}$ via the recoil curve.
Some of the hard ``down''-grains are switched back, 
whereas the weak grains shift their magnetization reversible.
The related mean magnetization is $\mmm{2}$.
The contribution of the weak grains results from eq. (\ref{mvonHsw}), if
we insert $\hex{2}$ for $\Hex$ and $\mmm{2}$ for $\mmm{}$ resp. followed by
an integration from zero to $nM_S/\mu_0$ with respect to $H_s$.
The contribution of the hard grains is more difficult to calculate, due to
the memory effect. Let us consider the probability $\wdu(H_s)$ that a hard 
grain with a given $H_s$ was
switched by $\hex{1}$ and switched back by $\hex{2}$ afterwards. 
To calculate this probability it is inevitable
to make an assumption on the correlation between the fluctuation of
the local field $H_1$ which acts on a grain if $\hex{1}$ is applied and the
fluctuation of $H_2$ according to the external field $\hex{2}$. Of course, if
$\hex{2}$ is only slightly different from $\hex{1}$ it is unlikely that the
neighbourhood of a grain changes considerably, so that $H_{1}$ and $H_{2}$
should be strongly correlated. With increasing distance between 
$\hex{1}$ and $\hex{2}$ this correlation will vanish, due to the multitude of
switching processes. The reasoning is as follows: The thermodynamical 
potential has surely a lot of nearly degenerated local minima.
Applying $\hex1$ selects one of them. A small change of the applied field 
is not enough to overcome the barrier between adjacent minima, but larger
changes generate energy surfaces, which are completely different in their 
topological structure, thus making states accessible, which may be very different
from the state at $\hex{1}$. The change
to such a state will be accompanied by a lot of correlated switchings. 
Even if the mean magnetization is changed only slightly, the neighbourhood
of an individual grain may be completely different, hence we can average with
respect to $H_1$ and $H_2$ independently, if the difference $\hex{2}-\hex{1}$
is not to small. Thus the probability $\wdu$ decouples accordingly:  
\beq
w_{\downarrow \uparrow}(H_s)&=& p_1(H_s) q_2(H_s)
\label{wdu}
\eeq
with $p_1(H_s)$ from eq. (\ref{p1}) and
\beq
q_{2}(H_s)&=& \int\limits_{-\infty}^{\infty} dH \, \romf_{(2)}(H)\,
                  \Theta(\hidown{2}-H_s) \, \Theta(\hiup{2}+H_s) \nonumber \\
          &=& \frac{1}{2} \, \Big ( 1-\romerf(y_L) \Big )
\eeq
with
\beq
y_L &=& \frac{\mu_0H_s - n M_s - n \mean{M}_2 - \mu_0\hex{2} 
+ N \mean{M}_2}{\mu_0\sigmaf \left ( 1-\mean{M}_2^2/M_s^2 \right )} \punkt
\eeq 
Again the switching conditions are included by help of related Heaviside
functions. Due to the mentioned magnetization changes in the environment
of a grain it may happen that grains which resisted $\hex{1}$ are switched
down by $\hex{2}$ (``up''-``down''-contribution). The related probability
factorizes also 
\beq
\wud(H_s)&=& \Big ( 1-p_1(H_s) \Big ) \, p_2(H_s)
\label{wud}
\eeq
with $p_1(H_s)$ again from eq. (\ref{p1}) and
\beq
p_{2}(H_s)&=& \int\limits_{-\infty}^{\infty}dH \,\romf_{(2)}(H)
                        \Theta(-H_s-\hiup{2})\,\Theta(H_s-\hidown{2}) \\
          &=& \frac{1}{2} \, \Big ( 1+\romerf(z_H) \Big ) 
\eeq
with
\beq
z_H &=& \frac{-\mu_0H_s + n M_s - n \mean{M}_2 - \mu_0\hex{2}
+ N \mean{M}_2}{\mu_0\sigmaf \left ( 1-\mean{M}_2^2/M_s^2\right )}
\punkt
\eeq
There are two further probabilities corresponding to the remaining two 
histories, which a grain may experience, i.e.
the probability that it resisted both $\hex{1}$ and $\hex{2}$ (
``up''-``up'' contribution) and the probability that a grain was switched
by $\hex{1}$ but resisted back-switching by $\hex{2}$ 
(``down''-``down'' contribution). We find
\beq
\wuu(H_s)&=& \Big ( 1-p_1(H_s)\Big ) \, \Big ( 1-p_2(H_s) \Big ) \label{wuu}
\komma
\\
\wdd(H_s)&=& p_1(H_s) \, \Big( 1-q_2(H_s) \Big ) \label{wdd}
\quad .
\eeq 
Thus hard grains with a given $H_s$ contribute to the magnetization
\beq
M_2(H_s)&=& M_s \Big ( \wuu(H_s) - \wud(H_s)
                     + \wdu(H_s) - \wuu(H_s) \Big ) 
\punkt
\eeq
Averaging with respect to $H_s$ yields the recoil curve
for the mean magnetization $\mmm{2}$ in state 2 in dependence on
both magnetic fields $\hex{1}$ and $\hex{2}$:
\beq
\mmm{2}&=&\int\limits_{0}^{nM_S/\mu_0}\,d\,H_s\,\romg(H_s)\, \mmm{w,2}(H_s) +
\int\limits_{nM_S/\mu_0}^{\infty}\,d\,H_s\,\romg(H_s)\, \mmm{2}(H_s) 
\label{recoilw}
\punkt
\eeq
In Fig. \ref{recoil} we show additional to the recoil curve a set of minor loops  
starting from different values of $\mean{M}_1$. 
\subsection{Calculation of the thermal remagnetization}
In the preceding sections we calculated the demagnetization and 
the recoil curves with the implicit understanding 
that the magnetization changes 
isothermally at a temperature $T_0$. The actual TR occurs if 
a dc-demagnetized RE-TM-magnet
is heated while $\hex{2}$ is kept constant (cf. Fig. \ref{prinzip}
from point ``2'' to point ``T''). 
Raising the temperature
has usually two effects. On the one hand the saturation magnetization decreases
with increasing temperature and on the other hand the switching fields $H_s$
are changed. Whereas the temperature dependence  of the saturation
magnetization can be measured easily, it is impossible to measure it for
the switching fields. 
What can be measured directly is the temperature dependence of the
coercive field. In appendix B we show that in our model $\bar{H}_s$ 
is related to $H_C$ and $M_S$ by the simple formula 
\beq
\mu_0\bar{H}_s(T)&=&\mu_0H_C(T)+nM_S(T)
\label{kronmuellerformel}
\eeq 
if the influence of the weak grains is negligible. 
Otherwise we have to solve
eq. (\ref{demagw}) for $\bar{H}_s$ with $\hex{1}=H_C$ and $\mean{M}_1=0$.
$H_C(T)$ and $M_S(T)$ are taken from measurements \cite{Jahn2000}.
Due to the lack of better information
we adopt the temperature dependence of $\bar{H}_s$ for arbitrary values of
the switching fields 
\beq
H_s(T)&=&H_s(T_0)\,\frac{\bar{H}_s(T)}{\bar{H}_s(T_0)} 
\label{HsvonT} \punkt
\eeq
The calculation of the magnetization 
$\mean{M}_2^T$ at an enhanced temperature $T$ is similar to the calculation
of the magnetization $\mean{M}_2$ by help of eq. (\ref{recoilw}), 
whereby the deformation of the switching field distribution 
has to be regarded. That should be done by indicating the
changed parameters with an index T, whereas the related parameters at the 
temperature $T_0$ will not be marked. 
From  eq. (\ref{HsvonT}) and the normalization of the switching
field distribution  follows that $\sigmas$ has the same temperature dependence
like $H_s(T)$.
The number of weak grains increases
with temperature, since $H_s(T)$ is usually much more reduced than $nM_S(T)$
if the temperature is raised. The magnetization of a weak grain with
given $H_s$ at the temperature T results from eq. (\ref{mvonHsw}) with
the appropriate $H_s(T)$, $M_S(T)$, and $\mean{M}_{2}^T$ inserted.
For the hard-grain contribution we have to calculate the probabilities
according to eqs. (\ref{wud},\ref{wdu},\ref{wuu},\ref{wdd}), 
however for the elevated temperature. 
We find for the magnetization of the hard-grain fraction with given $H_s$ 
\beq
M_2^T(H_s)&=& M_S^T  \Big ( \wuu^T(H_s) - \wud^T(H_s) 
                          + \wdu^T(H_s) - \wdd^T(H_s) \Big ) 
\eeq 
It is obvious that the probability
$p_1(H_s)$ remains unchanged, since the grains were switched downwards
at the initial temperature $T_0$. Therefore we have
\beq
\wuu^T(H_s)&=& \Big ( 1-p_1(H_s) \Big ) \, \Big ( 1-p_2^T(H_s) \Big )
\label{wuut}\\
\wud^T(H_s)&=& \Big ( 1-p_1(H_s) \Big ) \, p_2^T(H_s)  
\label{wudt}\\
\wdu^T(H_s)&=& p_1(H_s) \, q_2^T(H_s)  
\label{wdut}\\
\wdd^T(H_s)&=& p_1(H_s) \, \Big ( 1-q_2^T(H_s) \Big ) 
\label{wddt}
\eeq 
The probabilities 
\beq
q_{2}^T(H_s)&=& \frac{1}{2} \left ( 1-\romerf(y_L^T) \right )
\eeq
and 
\beq
p_{2}^T(H_s)&=& \frac{1}{2} \left ( 1+\romerf(z_H^T) \right ) 
\eeq
become temperature dependent both directly due to $H_s(T)$, $M_s(T)$, 
and $\sigmaf(T)$ and indirectly via $\mean{M}_2^T$, since we have 
now
\beq
y_L^T &=& \frac{\mu_0H_s^T - n M_S^T - n \mean{M}_2^T - \mu_0\hex{2} 
+ N \mean{M}_2^T}{\mu_0\sigmaf^T \left ( 1-(\mean{M}_2^T/M_S^T)^2 \right )} 
\eeq 
and
\beq
z_H^T &=& \frac{-\mu_0H_s^T + n M_S^T - n \mean{M}_2^T - \mu_0\hex{2}
+ N \mean{M}_2^T}{\mu_0\sigmaf^T \left ( 1-(\mean{M}_2^T/M_S^T)^2 \right )} \, .
\eeq
The temperature dependence of $\sigmaf^T=\sigmaf(T)$ was assumed to be that of $M_S$, 
since the fluctuations are caused by the inhomogeneities of the magnetization.
From the numerical solution of the implicit self-consistent equation
we calculated the TR in dependence on the parameters for the above mentioned
parameter set.
Here the starting temperature was fixed to \mbox{300 K} and we 
assumed a spherical sample. 
The influence of the 
switching field distribution width is studied in Fig. \ref{remagall}~A. 
To get noticeable TR-effects the distribution width $\sigmas$ should not be too small.
The dependence of the TR on the fluctuation width $\sigmaf$ is shown in
Fig. \ref{remagall}~B. 
The TR peak becomes broader and shifts to lower temperatures with increasing
field fluctuations.
The strong dependence of the TR on the internal demagnetization factor is 
depicted in Fig. \ref{remagall}~C.
For small demagnetization factors, i.e. for elongated grains, the TR 
is small, whereas effects up to 100 \% 
are possible for platelet-shaped grains.
Up to now there are  no experimental results available regarding
the influence of the internal demagnetization factor $n$ onto the TR. Regarding
the external demagnetization factor $N$ it
is known that in closed-circuit measurements TR-effects up to 100\% 
are possible \cite{Kavalerova75} (related to $M_S(T_{max})$ !) in the case
of SmCo$_5$ sintered magnets.
For samples measured in a VSM, where the demagnetization factor
is between 0.1 and 0.6, a significant lowering of the TR with increasing 
demagnetization factor N is reported. 
This is exactly what the calculated curves in 
Fig. \ref{remag_N} demonstrate. 
Another feature, which  has been studied experimentally, is
the dependence of the TR on the initial temperature $T_0$. 
The maximum TR increases 
and the peak position
shifts slightly to lower temperatures with decreasing $T_0$ \cite{Jahn2000}, 
what agrees well with our theoretical results depicted in Fig. \ref{remag_T0}.
\section{Discussion}
\label{discussion}
The above presented theory explains the observed features of the 
TR-ex\-peri\-ments well, whereby we focused especially to the shift of the 
temperature $T_{max}$, 
where the maximum TR occurs, as well as to the height of the maximum itself. 
Especially the strong dependence on the external demagnetization factor and on
the starting temperature can be explained. Furthermore the theory is able to 
explain TR effects of more than \mbox{50 \%}. 
To proceed to the description of the TR it was
inevitable to develop a theory for the hysteresis and the recoil curves. 
Whereas there exist a multitude of micromagnetic
calculations trying to explain the coercivity by assuming special 
microstructures which give rise to either pinning of domain walls or
nucleation, our theory starts a little above this level, since we do not
judge about the reason for a switching field, but simply accept its existence. 
In this point
our theory resembles the Preisach \cite{Preisach35} model, which 
was created to take into account the interaction between different grains
(bistable magnetic units).
The main disadvantage of the Preisach model is due to the fact that the 
Preisach distribution function is not dependent on the magnetization state 
\cite{Preisach35,Bertotti98}.
It is easy to prove that a theory of the TR based on this model is not able
to explain TR effects of more than 50\%. 
The reason is the lack of feedback.
Nevertheless, we share the opinion that the difference between the local fields 
and the mean field, which is due to grain interactions, is the crucial
point. In this work the width of these fluctuations is given by
$\sigma_f$. Our results demonstrate that without local field fluctuations
there will be no TR effect above 50 \%, e.g. in a normal mean-field theory.
To take into account the feedback of the neighbourhood of a grain
the inclusion approximation, albeit at a very rough level, was introduced. 
This provides us with a new parameter $n$, i.e. the internal demagnetization
factor. 
Assigning the same internal demagnetization factor $n$ to all grains  
seems to be a severe simplification, but since we have no reliable
information on the switching field distribution, we probably account, at least
partly, for different grain shapes and/or exchange coupling via the grain
boundaries by adjusting the switching field
distribution. Thus, $n$ is more a feedback parameter than a real
demagnetization factor.
The most difficult problem is the
determination of the temperature dependence of the switching fields. All the
former theories used the temperature dependence of the coercivity for this task, 
due to the lack of better information. But it is evident
that this forces the TR magnitude to drop to zero at the temperature T$_{H_C=0}$
where the coercivity vanishes. 
Contrary, investigations on SmCo$_5$ demonstrated
that the TR manifestly remains above T$_{H_C=0}$ \cite{Jahn2000}.
This is due to the fact that for this compound the difference
between T$_{H_C=0}$ and the Curie temperature $T_C$ 
is about 250 K \cite{Adler85}. 
Since our theory relates the mean switching field to the coercivity, it
was possible to extract the temperature dependence $\bar{H}_s(T)$ 
from the measured $H_C(T)$ curves. The derived
formula is in congruence with the empirical formula used in 
\cite{Hirosawa86,Sagawa88} if one relates $cH_A$ (c$\ll 1$) 
to the switching field and
$n$ to the corresponding effective demagnetization factor $N_{eff}$. 
This holds as long
as the mean switching field is greater than the internal demagnetizing field. 
Otherwise the simple rule given in eq. (\ref{kronmuellerformel}) fails.
Another interesting point is whether the weak grains are essential for
the description of the TR. 
The calculated relative volume 
fraction of weak grains in dependence on the temperature is
depicted in a dashed fashion in Fig. \ref{probabilities}. 
At $T_{max}$ the   
$\bar{H}_s$ equals roughly $nM_S/\mu_0$, i.e. the maximum TR is observed
shortly before the majority of the grains become weak. 
At temperatures higher than $T_{max}$ the magnetization decreases. 
Sm-Co-magnets undergo irreversible changes in
their microstructure at higher temperatures. That is why measurements 
above $T_{max}$ are rather difficult and, therefore, seldom reported.
For the parameter set used in Fig. \ref{approximations} the influence 
of weak grains
is small as long as the temperature is below $T_{max}$. For $T>T_{max}$ 
the neglection of the weak grains would result in a
qualitative different behaviour, since we get a sign change of the TR
before it finally vanishes, as shown in Fig. \ref{approximations} in 
dotted manner.
The reason for that peculiarity is made clear in Fig.
\ref{probabilities}, where we have plotted the temperature dependence of the
probabilities $\wuu$, $\wud$, $\wdu$, and $\wdd$ regarding the hard grain
fraction. 
It is obvious that at first
$\wdu$ increases and when it drops down $\wud$ rises up, thus
explaining the sign change. This behaviour is not observed if the probability 
$p_2$, i.e. that a grain which resisted $\hex{1}$ will be switched by $\hex{2}$,
is neglected, as it was done in former theories, but then an overestimation of
the TR occurs, as visible in Fig. \ref{approximations}. For the description
of the TR at higher temperature it is necessary to take into account, 
that above $T_{max}$ even the switching fields of the hardest grains are
lowered enough to be switched by the still existing stray fields. The latter 
do not vanish until the magnetization breaks down. 
It is worth to point out, that weak grains are not necessary to get effects of 
more than 50 \%. 
Such a large TR one gets due to the feedback of the magnetization via the
internal demagnetizing fields. In some sense this models  
the well known avalanche effects observed during hysteresis measurement, 
which should also occur to a certain extent if the sample remagnetizes.
In consideration of the various approximations one can not expect that all magnetic
measurements are explained quantitatively by the presented theory. 
Nevertheless in Fig. \ref{remagfit}, where we present a fit of the theory to our
measurement of the TR in a SmCo$_5$ sample (VACOMAX 200), 
the agreement is rather well, especially in comparison with Fig. 2 in
ref. \cite{Lileev92}.   
If one calculates the hysteresis curve with the parameters fitted to the TR
curve the agreement is not as well as for the TR.
This is not astonishing, since the theory was designed mainly for
the qualitative understanding of the TR, which is very sensitive to
the field fluctuations but less sensitive to the switching field distribution.
For the hysteresis the situation is vice versa. 
To get a good description of
the hysteresis the $H_s$ spectrum has to be determined more accurately.
Another reason may be the simplification in the description of the grain 
interaction. Correlated switching of a multitude of grains is taken into account
in a simplified manner via $n$ only, as discussed above.
Surely this is more important for the hysteresis than for the TR.
The neglection of the angular fluctuations of the field
and the reversible rotation processes are in this regard of less 
importance, since this will influence both the
hysteresis and the TR similarly.
Since the present paper is focused on the theory, we save the detailed 
analysis of experimental results, as well as the modifications necessary
to describe multi-phase magnets and the inverse TR in Ba-ferrites 
to forthcoming papers.  
In conclusion we may say that the presented theory 
is able to describe the rather 
complex findings by help of only three
internal parameters, i.e. $\sigmas$, $\sigmaf$, and $n$, 
which may be helpful in characterizing
different hard magnetic materials 
more effectively.  
\section*{Acknowledgement}
We would like to thank V. Ivanov and  K.H. M\"uller for useful discussions.
One of us (L.J.) acknowledges the support by the Deutsche
Forschungsgemeinschaft, Grant LO 293/1-1.
\begin{appendix}
\section*{Appendices}
\section{The magnetization of a weak grain}
We want to find the mean internal magnetization of a grain in a MDS.
For that purpose we assume, that the grain contains a lot of domains
separated by 180$^{\circ}$-domain walls. 
The dimension in field direction is large with respect to the dimension
in a perpendicular direction. Thus the demagnetizing factor of such a
slap-shaped domain is approximately zero. 
The average internal magnetization is then
\beq
\mean{M}_i&=&(1-\lambda)M_S - \lambda M_S  \label{Minnen}
\eeq
with $\lambda$ being the relative volume of the downwards magnetized domains.
The magnetic energy density contains both  
the Zeeman energy
\beq
w_H&=&-\mean{M}_i H
\eeq
and the energy of the stray fields  
\beq
w_{stray}&=& -\int H_{dem}\, d\mean{M}_i  \label{wstreu}
\eeq

Substituting for the demagnetization field
\beq
\mu_0H_{dem}&=& -n \,(\mean{M}_i-\mean{M})
\eeq
an integration of eq. (\ref{wstreu}) yields 
the total energy density $w_{tot}$ in dependence on $\lambda$ to be
\beq
w_{tot}&=& - \frac{M_s^2}{\mu_0}
\left ( \frac{\mu_0 H}{M_S}+n\frac{\mmm{}}{M_S} \right )\,(1-2\lambda)
+\frac{nM_s^2}{2\mu_0} \,(1-2\lambda)^2
\eeq
If the $\lambda$, which minimizes the energy, is inserted into
eq. (\ref{Minnen}) one gets eq. (\ref{weakmagnetization}),
as long as $\lambda$ is between zero and one. Otherwise the grain is saturated
in forward or backward direction.
Thus the MDS grains show no memory effect, but follow the local field
immediately.
\section{Relation between ${\rm \bf H_C}$ and ${\rm \bf \bar{H}_s}$}
The influence of the weak grains is negligible, 
if $\mu_0\bar{H}_s\gg nM_S$ holds and the
distribution width $\sigmas$ is small enough. The first integral in
eq. (\ref{demagw}) may be neglected and in the second one we can extend the lower
integration limit to minus infinity. If the external field $\hex{1}$ equals
$-H_C$ we have $\mmm{1}=0$ simultanously. Therefore we find now
\beq
0&=&-\int\limits_{-\infty}^{\infty}\,d\,H_s\,\romg(H_s)\, \romerf(x_H)
\label{demagw0}
\eeq
Substituting $x=(H_s-\bar{H}_s)/\sigmas$ the switching field distribution 
becomes symmetrical with respect to $x=0$, thus the monotonous 
function $\romerf(x_H)$ has to be asymmetrical to reduce the integral to zero.
Otherwise we find from eq. (\ref{p1}) that the $\romerf$-function is shifted, 
since we have $x_H=(-\bar{H}_s-\sigmas x+nM_S/\mu_0+H_C)/\sigmaf$. To make the
erf-function asymmetrical, the shift has to be zero. 
This yields eq. (\ref{kronmuellerformel}). 
\end{appendix}
%
%






\newpage
\section*{Captions of figures}
\begin{figure}[!h] 
\caption{ 
\label{abb:prinzip} Scheme of the TR-experiment.
From the saturated state the sample is isothermally driven to point ``1''
and afterwards along the recoil curve to point ``2''.
The TR happens, if the sample is 
heated (point ``2'' to point ``T''), while the external field is 
kept constant (enhanced insert on the
right hand side). For the external fields at point ``1'' and ``2'' the
sheared remanence coercivity $H_R^{ext}$ and $\Hex=0$ zero resp. were chosen, since the most 
TR-experiments start from the dc-demagnetized state.
}
\label{prinzip}
\end{figure} 
\begin{figure}[!h] 
\caption{ 
\label{cap:inclusion} The grain is considered as an inclusion with
magnetization $\pm M_S$ (here we assumed $-M_S$, symbolized by the long white
arrow) embedded in an environment differing in its magnetic field $H$ from the mean magnetic
field $\mean{H}$  within the sample due to the fluctuation $\Delta H$. 
Fluctuations of the magnetization in the environment are neglected, thus the
magnetization in the environment is equal to the mean magnetization of
the sample. This is symbolized by the two shorter white arrows of equal length.
}
\label{inclusion}
\end{figure} 
\begin{figure}[!h] 
\caption{ 
\label{abb:recoil} The demagnetization curve, the recoil curve, 
and a set of minor loops starting from  different values of $\mean{M}_1$ 
calculated for a representative parameter set close to $SmCo_5$-mangets at 300K. 
Both the sample and the grains are adopted as spheres.
}
\label{recoil}
\end{figure} 
\begin{figure}[!h] 
\caption{ 
\label{abb:remagall} The TR 
in dependence on (A) the width of the switching field distribution $\sigmas$,
(B)  on the width of field fluctuations $\sigmaf$, and
(C) on the internal demagnetization factor $n$.
}
\label{remagall}
\end{figure} 
\begin{figure}[!h] 
\caption{ 
\label{abb:remag_N} The TR 
for different external demagnetization factors $N$. For simplicity this
figure was calculated with $\bar{H}_s(T)=H_C(T)$.
}
\label{remag_N}
\end{figure} 
\begin{figure}[!h] 
\caption{ 
\label{abb:remag_T0} The TR 
for different values of the initial temperature $T_0$ (parameters on the curves).
} 
\label{remag_T0}
\end{figure} 
\begin{figure}[!h] 
\caption{ 
\label{abb:approximations} The TR for different levels of
approximation. The solid line gives the results of the presented theory. 
The dot-dashed line
results from neglecting the weak grain fraction and for the calculation
of the dashed line, both the weak grains and $p_2^T$, i.e. the probability,
that a grain which resisted $\hex{1}$ will be switched while heating at
$\hex{2}=0$, were neglected.
}
\label{approximations}
\end{figure} 
\begin{figure}[!h] 
\caption{ 
\label{abb:probabilities} The probabilities $\wuu^T$, 
$\wud^T$, $\wdu^T$, and $\wdd^T$ (cf. eqs. (\ref{wuut}-\ref{wddt}))
in dependence on the temperature. The dashed line shows the
ratio of the volume of weak grains to the total volume.
\label{probabilities}
}
\end{figure} 
\begin{figure}[!h] 
\caption{ 
\label{abb:remagfit} The TR and H$_C$(T) measured for a SmCo$_5$
sample (VACOMAX 200) together with the calculated TR, where the parameters
$\sigmas$, $\sigmaf$, and $n$ have been adjusted accordingly.
}
\label{remagfit}
\end{figure}
%
%
\begin{center}
\newpage
\thispagestyle{empty}
\begin{minipage}[t]{12cm}
\psfig{file=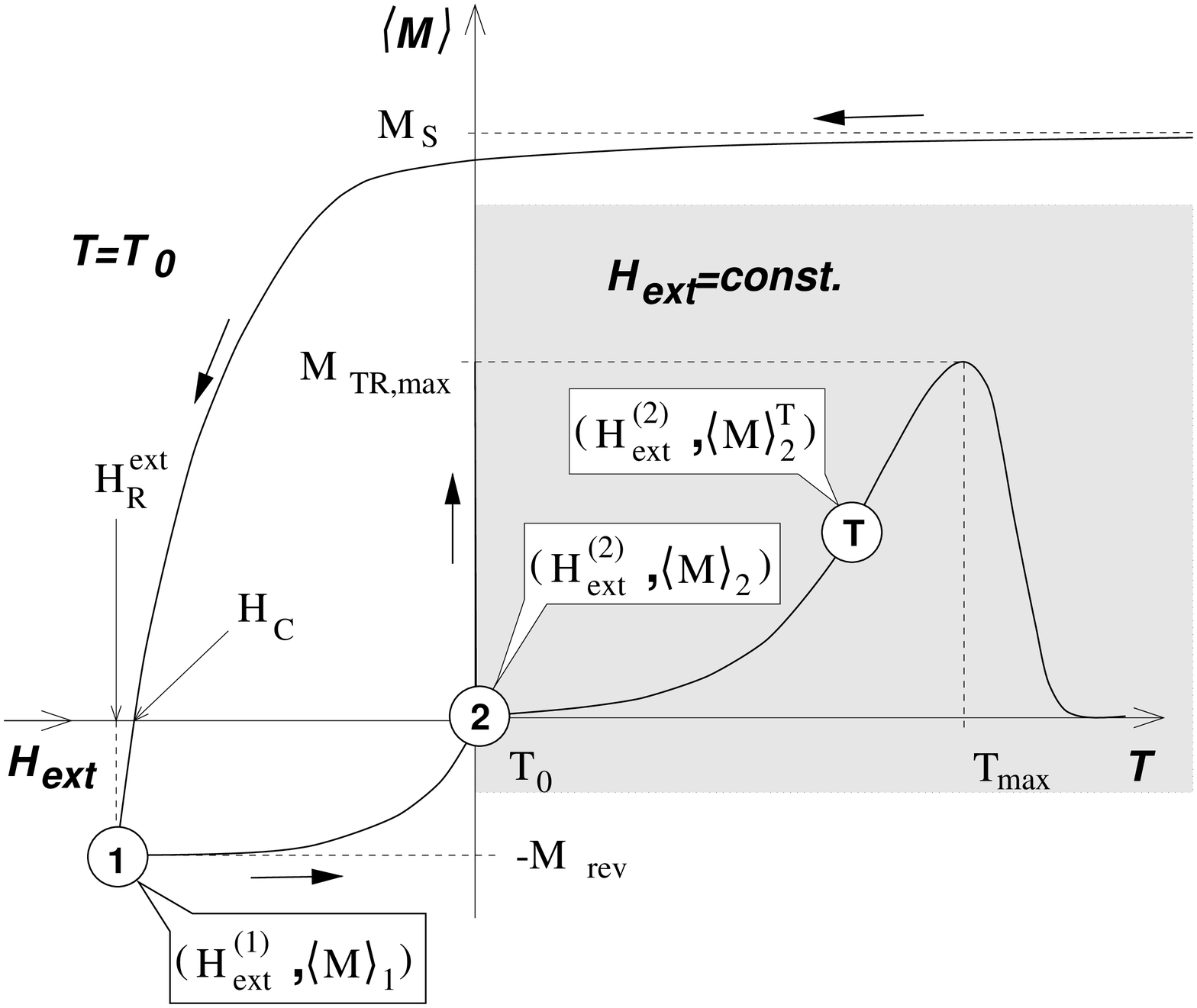,width=12cm}
\end{minipage}\\ \vspace*{\fill} Figure 1
\newpage
\thispagestyle{empty}
\begin{minipage}[t]{10cm}
\psfig{file=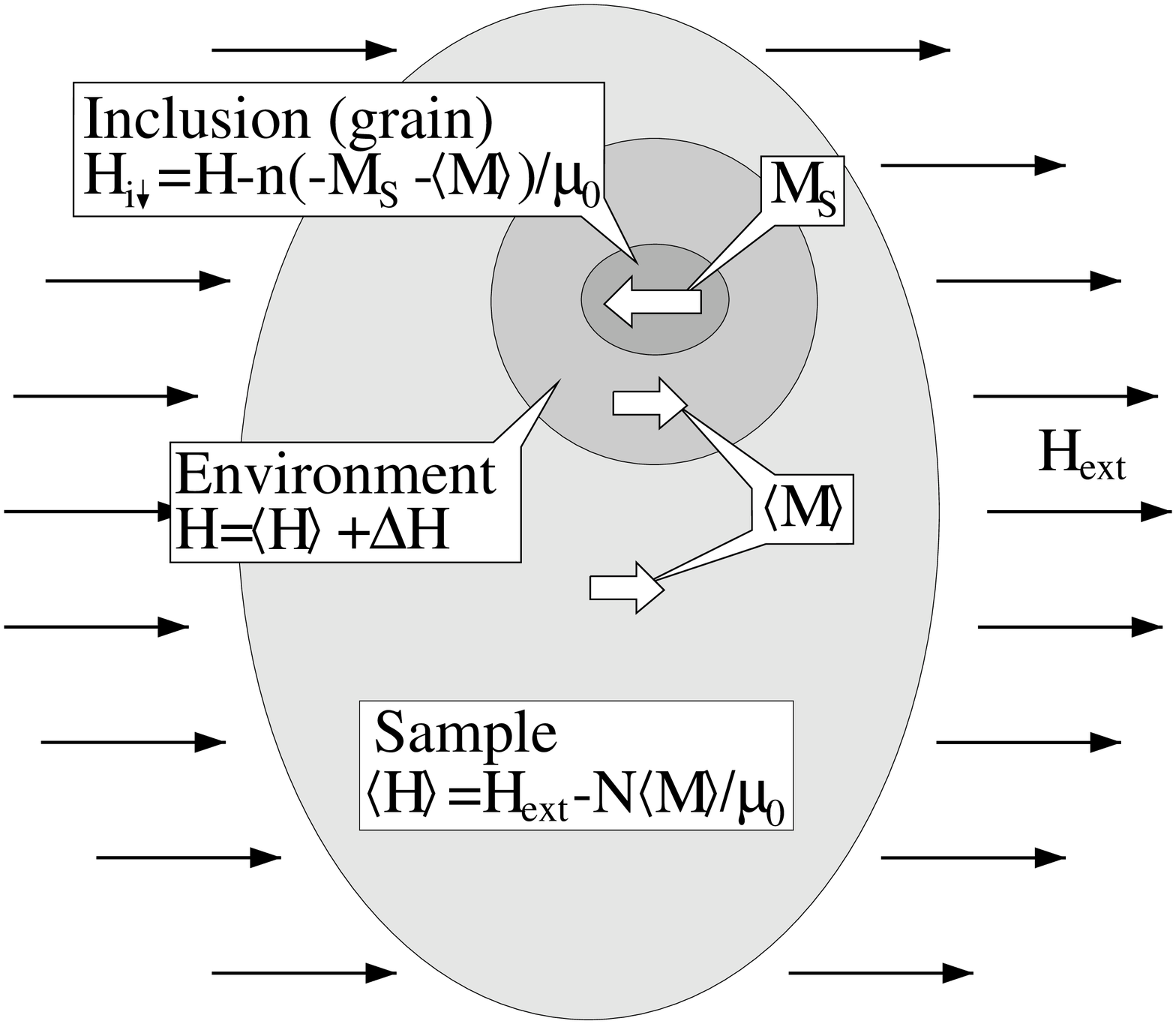,width=12cm}
\end{minipage}\\ \vspace*{\fill} Figure 2
\newpage
\thispagestyle{empty}
\begin{minipage}[t]{12cm}
\psfig{file=recoil.eps,width=12cm}
\end{minipage}\\ \vspace*{\fill} Figure 3
\newpage\thispagestyle{empty}
\begin{minipage}[t]{10cm}
\psfig{file=remagall.eps,width=10cm}
\end{minipage}\\ \vspace*{\fill} Figure 4
\newpage\thispagestyle{empty}
\begin{minipage}[t]{12cm}
\psfig{file=remag_N.eps,width=12cm}
\end{minipage}\\ \vspace*{\fill} Figure 5
\newpage\thispagestyle{empty}
\begin{minipage}[t]{12cm}
\psfig{file=remag_T0.eps,width=12cm}
\end{minipage}\\ \vspace*{\fill} Figure 6
\newpage\thispagestyle{empty}
\begin{minipage}[t]{12cm}
\psfig{file=approximations.eps,width=12cm}
\end{minipage}\\ \vspace*{\fill} Figure 7
\newpage\thispagestyle{empty}
\begin{minipage}[t]{12cm}
\psfig{file=probabilities.eps,width=12cm}
\end{minipage}\\ \vspace*{\fill} Figure 8
\newpage\thispagestyle{empty}
\begin{minipage}[t]{12cm}
\psfig{file=remagfit.eps,width=12cm}
\end{minipage}\\ \vspace*{\fill} Figure 9
\end{center}
\vspace*{\fill}
\end{document}